\documentclass{PoS}

\usepackage{epsfig}
    
\PoS{PoS(LAT2005)177}

\title{The screening length in hot QCD}

\ShortTitle{The screening length in hot QCD}

\author{Olaf Kaczmarek\\
  Fakult\"at f\"ur Physik, Bielefeld University, D-3361 Bielefeld (Germany)\\
  E-mail: \email{okacz@physik.uni-bielefeld.de}}

\author{\speaker{Felix Zantow}\\
  Brookhaven National Laboratory, Upton, NY-11973 (USA)\\
  E-mail: \email{zantow@quark.phy.bnl.gov}}

\abstract{We discuss the temperature dependence of the screening lengths in
  quenched and full QCD using the non-perturbative lattice approach. We analyze
  the temperature dependence of distances which are defined as moments of quark
  antiquark free energies, {\em i.e.} we introduce
\begin{eqnarray}
\langle r^n\rangle &\equiv\frac{1}{{\cal Z(T)}}\int d^3r\; r^n F(r,T)\;,\qquad (n=1,2,3,...)\nonumber
\end{eqnarray}
where ${\cal Z}(T)$ is a suitably chosen normalization constant and $F(r,T)$
denotes the change in free energy due to the presence of heavy quarks. These
distance scales are supposed to describe the geometric size of partonic clouds
which screen static charges in a medium and characterize distances beyond which
the quark antiquark free energy is to large extent dominated by medium effects.
At asymptotic high temperatures these moments can be related to the inverse
Debye mass. In our numeric analysis we find that these moments drop rapidly in
the vicinity of the phase transition and indicate distances which are about
twice as large as the inverse Debye mass found in earlier studies. Our analysis
supports recent findings that indicate that $J/\psi$ will show significant
medium modifications only at temperatures well above $T_c$.}

\FullConference{XXIIIrd International Symposium on Lattice Field Theory\\

  25-30 July 2005\\
  
  Trinity College, Dublin, Ireland}

\begin{document}

\section{Introduction}
In a hot and dense medium the screening radius rapidly decreases with
increasing temperature and/or density and the occurrence of the phase
transition in QCD towards a deconfined medium is intimately connected to a
decreasing screening length. Due to unambiguities in perturbation theory at
temperatures only moderately above phase transition most of todays quantitative
discussions of the screening mass and length, and corresponding medium effects
on heavy quark bound states \cite{Matsui:1986dk,Karsch:2005ex}, are to large
extent based on the non-perturbative lattice approach. Here, the Debye
screening mass, $m_D(T)$, has been extracted from either the infrared limit of
the gluon propagator, which should give a proper definition of the Debye mass,
or from the screened Coulomb behavior of quark antiquark free energies at large
distances. In such lattice studies large values for the screening mass were
obtained
\cite{Heller:1997nq,Kaczmarek:1999mm,Digal:2003jc,Nakamura:2004wr,Kaczmarek:2004gv,Kaczmarek:2005ui}
which indicate a quite small screening length,
\begin{eqnarray}
r_D(T)&\equiv&\frac{1}{m_D(T)}\;.\label{eq1}
\end{eqnarray} 
We will introduce here a screening length by characterizing the temperature
dependence of the geometric size of gluonic clouds that are induced due to the
presence of static color charges in the medium. For this purpose we introduce
and discuss moments of distance dependent quark antiquark (free) energies at
finite temperatures.
 
\section{Quark antiquark interactions and screening}\label{Sec2}
\begin{center}
\begin{figure}[t]
\begin{center}
  \epsfig{file=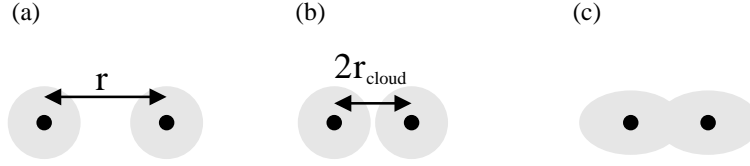,width=10cm}
\end{center}
\caption{Illustration of screening through the polarization of gluons in the medium: (a) At large distances ($r\to\infty$) the gluon clouds which surround the static test charges are well separated. At smaller distances, $r<2r_{cloud}$, the clouds begin to overlap (b) and the geometric structure of the clouds will depend on the separation between the test charges (c).}
\label{fig1}
\end{figure}
\end{center}

In a heat bath at fixed temperature the parton density distribution can be
considered as being homogenous and the mean free path of partons is
determined only by temperature and (parton) density. The presence of static
charges in the medium, however, will polarize the medium and the parton density
distribution will change compared to the density distribution of the heat bath
without static charges. In particular, the parton density is expected to
increase in the vicinity of static test charges and will depend also on the
distance between them. This property is illustrated in Fig.~\ref{fig1} for the
case of a static quark antiquark ($Q\bar Q$) pair: The region at which one
expects a significant higher parton density might be well located in the
vicinity of the test charges and will be considered by us in the following as
{\em clouds} which screen and neutralize the additional color charges. At
sufficiently large separation of the $Q\bar Q$ pair ($r\to\infty$) the
interaction is screened through these clouds and both clouds will be well
separated (see Fig.~\ref{fig1}~(a)). In this case the energy (and entropy)
which is needed to neutralize the charges will not depend on distance. When
going to smaller distances, however, the clouds begin to overlap (b) and the
geometric structure of the density distribution is supposed to depend also on
distance (c). In this case the interaction energy of the quark antiquark will
show also significant $r$-dependence.

The distance below which the interaction is significantly modified by distance
could thus be used to characterize the sizes of the separated screening clouds.
To get a more quantitative understanding of the screening length and the size
of the clouds that surround the static quarks, we consider momenta, $\langle
r^n\rangle$, which are supposed to characterize the parton densities induced by
the presence of strong interactions, {\em i.e.} we consider
\begin{eqnarray}
\langle r^n\rangle_{1,8,av}&=&\frac{4\pi}{{\cal Z}_{1,8,av}(T)}\int_0^{\infty} dr\;r^{n+2}\;F_{1,8,av}(r,T)\;,\label{r}
\end{eqnarray}
where the temperature dependence of the normalization constants is given by
\begin{eqnarray}
{\cal Z}_{1,8,av}(T)&\equiv&4\pi\int_0^\infty dr r^2\;F_{1,8,av}(r,T)\;.\label{norm}
\end{eqnarray}
In both relations we already have assumed that the free energies are normalized
such that they approach zero at large distances, {\em i.e.}
$F_{1,8,av}(r\to\infty,T)=0$. In particular, when assuming a color screened
Coulomb behavior for $F_{1,8}(r,T)$ at high temperature and large distances,
{\em i.e.}
\begin{eqnarray}
F_{1,8}(r,T)=-C_{1,8}\alpha(T) exp(-m_D(T) r)/r\;,\label{fit}
\end{eqnarray}
(with $C_1=4/3$ and $C_8=-1/6$ for three colors, $N=3$) we obtain
\begin{eqnarray}
\langle r^n\rangle_{1,8}&=&(n+1)!\left(\frac{1}{m_D(T)}\right)^n\;.
\end{eqnarray}
To lowest order in $g$, thus, the size of the screening cloud is not expected
to depend on the specific color representation of the $Q\bar Q$ state (singlet,
octet) and the lowest moments, $\langle r\rangle_{1,8}$, are expected to be
about twice as large as indicated by the inverse Debye mass\footnote{Of course,
  at small distances the coupling will be promoted to a coupling which runs
  also with distance, $\alpha(r,T)$, and thus the relations deduced here (and
  in what follows) give only the leading approximation for the relations
  between the Debye mass, $m_D$, and the given momenta at high temperature},
\begin{eqnarray}
\langle r\rangle_{1,8}\;=\;2\frac{1}{m_D}\;=\;2r_D\;\;.\label{relrmda}
\end{eqnarray}  
Of course, similar relations can be obtained also for the manifest gauge
invariant observables $\langle r^n\rangle_{av}$ using the high temperature
perturbative relation $F_{av}\simeq-(F_1/T)^2/16$. In this case smaller
distances are obtained, {\em i.e.} $\langle r^n\rangle_{av}=n!/(2m_D)^n$. In
the limit of high temperatures the perturbative leading order relation between
the coupling and the Debye mass indicate only a slowly decreasing screening
length with increasing temperatures, {\em i.e.} $\langle
r\rangle_{1,8,av}\sim1/(gT)$.

On the other hand, at temperatures close but above $T_c$ remnants of the
confinement forces contribute to the non-perturbative properties of the quark
antiquark free energies \cite{Kaczmarek:2004gv,Kaczmarek:2005ui} and deviations
from a color screened Coulomb behavior become thus important already at
intermediate distances. In particular, assuming a string-type potential in QCD
at $T=0$,
\begin{eqnarray}
V(r)\simeq\left\{\begin{array}{l@{\quad:\quad}l}
-\frac{\pi}{12r}+\sigma(r-r_{\rm{break}})+\frac{\pi}{12r_{\rm{break}}} & r\;<\;r_{\rm{break}}\\ 
0 & r\;>\;r_{\rm{break}}
\end{array}\;,\right.
\end{eqnarray}
where $\sigma\simeq4.5$ fm$^{-2}$ denotes the string tension and
$r_{\rm{break}}\simeq(1.2\;-\;1.4)$ fm the distance at which the string breaks
at $T=0$, yields $\langle r\rangle_{T=0}\simeq(0.71\;-\;0.83)$ fm at $T=0$.

\section{Numeric results for screening mass and length}\label{Sec3}
Our numeric studies of the length scale $\langle r^n\rangle_{1,8,av}$ in QCD
are to large extent based on lattice calculations of quark antiquark free
energies given in Refs.~\cite{Kaczeetal,Kaczmarek:2005ui} for QCD and
calculations of free energies in quenched QCD
\cite{Kaczmarek:2002mc,Kaczmarek:2004gv}. We also compare our results to the
recent non-perturbative analysis of the Debye mass given in
Refs.~\cite{Kaczmarek:2004gv,Kaczmarek:2005ui} for $SU(3)$ and $2,3$-flavor QCD
and for $SU(2)$ \cite{Heller:1997nq}. To convert these masses to physical units
we used $T_c\simeq270$ MeV for quenched and $T_c\simeq203$ MeV in $2$- and
$T_c\simeq195$ MeV for $3$-flavor QCD, which are the relevant transition
temperatures at the quark mass values used in these calculations. We fixed the
transition temperatures for $SU(2)$ being 290 MeV.
\begin{figure}[tbp]
\begin{center}
  \epsfig{file=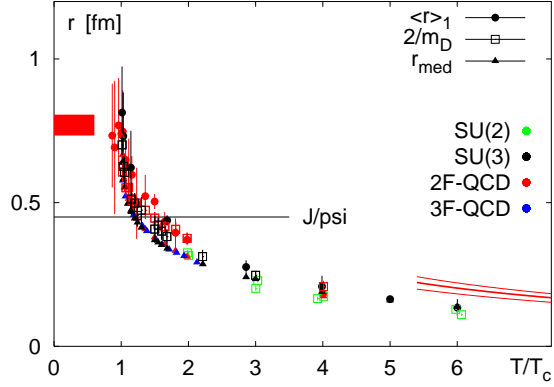,width=8.0cm}
\end{center}
\caption{Lattice results for $\langle r\rangle_1$ as function of $T/T_c$ in quenched ($N_f=0$) and full QCD ($N_f=2$) and compared to the size of $J/\psi$ at $T=0$. We also compare here $\langle r\rangle_1$ to different length scales obtained in terms of $2/m_D$ in $SU(2)$ \cite{Heller:1997nq}, $SU(3)$ \cite{Kaczmarek:2004gv} and $2$-flavor QCD \cite{Kaczmarek:2005ui} as well as to distances obtained for $SU(3)$ \cite{Kaczmarek:2002mc} and $2$- and $3$-flavor QCD \cite{Kaczmarek:2005ui} in terms of $r_{med}$. Further details on this figure, in particular on the different lines indicated here, are given in the text (see Sec.~3). }
\label{RES1}
\end{figure}

In Fig.~\ref{RES1} we summarize our lattice results for $\langle r\rangle_{1}$
obtained in quenched and full QCD (filled red circles: $2$-flavor QCD; filled
black circles: quenched QCD) in physical units as function of $T/T_c$. It can
be seen that $\langle r\rangle_1\simeq0.7$ fm at temperatures in the close
vicinity of the transition while $\langle r\rangle_1$ drops rapidly to values
below $0.4$ fm already at temperatures above $2T_c$. At higher temperatures
$\langle r\rangle_1$ continues to decrease rather slowly with increasing
temperatures. In particular, in the high temperature limit $\langle r\rangle_1$
is expected to decrease proportional to $(gT)^{-1}$. This behavior is also
indicated for $2$-flavor QCD in Fig.~\ref{RES1} by the red lines using
Eq.~\ref{relrmda} and the temperature dependence of the perturbative $2$-loop
coupling in the $\overline{MS}$-scheme (using renormalization scales
$\mu\in[\pi,4\pi]$). At the highest temperatures analyzed by us this estimate
still leads to larger distances. It is, however, quite interesting to note here
that at high temperatures similar distances to those indicated through $\langle
r\rangle_1$ are also indicated through twice the inverse Debye mass, $2/m_D$.
These distances are also shown in Fig.~\ref{RES1} by triangles.

Following Ref.~\cite{Matsui:1986dk} the decreasing behavior of (twice) the
screening length above $T_c$ may imply the onset of medium modifications on
heavy quark bound states. To estimate the relevant temperatures at which medium
effects on $J/\psi$ may become important we also indicate in Fig.~\ref{RES1}
the size of this state at $T=0$ by a horizontal line. At appears quite
reasonable that beyond temperatures at which twice the screening length becomes
smaller than the size of the particular bound state screening effects may
strongly influence the binding properties. From Fig.~\ref{RES1} it may thus
appear reasonable that $J/\psi$ survives the transition and may show
significant medium modifications at temperatures only well above $T_c$. This
can also be seen when comparing the data for $\langle r\rangle_1$ and $2/m_D$
in Fig.~\ref{RES1} with the scale $r_{\rm{med}}$ defined in
\cite{Kaczmarek:2002mc}, which is also shown in Fig.~\ref{RES1} by triangles.
In fact, $r_{\rm{med}}$ has been introduced to indicate distances at which
medium effects become important in the singlet free energy and both scales,
$\langle r\rangle_1$ and $r_{\rm{med}}$, indicate similar distances above $T_c$
in the entire temperature range shown in Fig.~\ref{RES1}. In particular, when
comparing the different scales discussed above for quenched and $2$-flavor QCD
with todays attainable scales for $SU(2)$ and $3$-flavor QCD no or only little
differences appear at temperatures only moderately above the transition. Of
course, the screening length in $SU(2)$ will diverge when going to temperatures
in the close vicinity of $T_c$.

Finally we also compare in Fig.~\ref{RES1} the values for $\langle r\rangle_1$
obtained in quenched and full QCD at low temperatures to the value one may
expect at zero temperature, $\langle r\rangle_{T=0}$, using the
parameterization for $V(r)$ discussed in Sec.~\ref{Sec2}. This value is
indicated by a thick line (left-hand side in Fig.~\ref{RES1}). It can be seen
that the values obtained for $\langle r\rangle_1$ become indeed quite similar
to $\langle r\rangle_{T=0}\simeq0.74\;-\;0.85$ fm already at temperatures close
but above $T_c$. Again this may demonstrate the presence of remnants of the
confinement forces above the transition and indicates that the singlet free
energy deviates at intermediate and small distances from the leading order
color screened Coulomb behavior expected at high temperatures. However, the
rather small differences observed between $2/m_D$ and $\langle r\rangle_1$ at
low temperatures presumably implies that the scale $\langle r\rangle_1$ is not
very sensitive to the small distance properties of $F_1(r,T)$.


\section{Conclusions}\label{Sec4}
We have discussed the screening length in quenched and full QCD using
properties of quark antiquark free energies calculated on the lattice. For this
purpose we analyzed moments of the free energies, {\em i.e.} $\langle
r\rangle_{1,8,av}$ defined in Eq.~(\ref{r}), which are supposed to characterize
distances at which medium modification become important in quark antiquark free
energies. We find that $\langle r\rangle_1$ drops rapidly at temperatures close
above the phase transition while at high temperatures $\langle r\rangle_1$ is
almost twice as large as the scale one obtains from the inverse Debye mass. A
comparison of $\langle r\rangle_1$ with the size of $J/\psi$ at $T=0$ suggests
that above temperatures about $1.3T_c$ the binding properties of $J/\psi$ could
be strongly influenced by screening. In particular, the length scales obtained
for $SU(3)$ as well as for QCD ($N_f=2,3$) show no or only little differences
at temperatures which are moderately above the transition. Our analysis of the
screening length in QCD thus supports recent lattice studies of quarkonium
spectral functions above $T_c$ in quenched \cite{Asakawa:2003re,Datta:2003ww}
and full QCD \cite{Skullerud} which indicate that $J/\psi$ may survive the
transition and will dissolve at higher temperatures. At low temperatures,
however, a more detailed analysis of $\langle r\rangle_{1,8,av}$ including also
higher moments, will be interesting in future. A similar analysis of the
screening length with this method may also become of considerable interest in
QCD at finite temperatures and densities \cite{Doering} using mesonic or
baryonic \cite{Hubner:2004qf} heavy quark free and/or internal energies
\cite{Kaczmarek:2005gi}.

\subsection*{Acknowledgments}
We thank F. Karsch, E. Laermann and H. Satz for helpful discussions. This work
has partly been supported by DFG under Grant No. FOR 339/2-1 and by BMBF under
Grant No.06BI102 and partly by Contract No. DE-AC02-98CH10886 with the U.S.
Department of Energy.

\end{document}